\documentclass[a4paper,11pt]{article}
\usepackage{pos}

\title{Multi-point Padè for the study of phase transitions: from the Ising model to lattice QCD}
\ShortTitle{Multi-point Padè for the study of phase transitions}

\author*{Francesco Di Renzo}
\author{Simran Singh}

\affiliation[]{Dipartimento di Scienze Matematiche, Fisiche e
  Informatiche, Università di Parma\\ 
and INFN, Gruppo Collegato di Parma, I-43100, Parma, Italy}

\emailAdd{francesco.direnzo@unipr.it}
\emailAdd{simran.singh@unipr.it}

\abstract{The Bielefeld Parma collaboration has recently put forward a
  method to investigate the QCD phase diagram based on the computation
  of Taylor series coefficients at both zero and imaginary values of
  the baryonic chemical potential. The method is based on the
  computation of multi-point Pad\'e approximants. We review the
  methodological aspects of the computation and, in order to gain
  confidence in the approach, we report on the application of the
  method to the two-dimensional Ising model (probably the most popular
  arena for testing tools in the study of phase transitions). Besides
  showing the effectiveness of the  multi-point Pad\'e approach, we
  discuss what these results can suggest in view of further progress
  in the study of the QCD phase diagram. We finally report on very
  preliminary results in which we look for Pad\'e approximants at
  different temperatures and fixed values of the (imaginary) baryonic 
  chemical potential.}

\FullConference{%
  The 39th International Symposium on Lattice Field Theory (Lattice2022),\\
  8-13 August, 2022 \\
  Bonn, Germany 
}

\begin{document}
\maketitle

\section{How it all began: from Taylor expansions on thimbles to
  imaginary $\mu_B$ LQCD}
\label{sec:PADEthimblePR}
The QCD phase diagram is still to a large extent elusive: in particular, due to the
so-called sign problem, the lattice (the non-perturbative tool which
would be supposed to provide valuable insight) cannot probe the
relevant regions in the $T-\mu_B$ (Temperature-baryonic chemical
potential) plane. In the last couple of years, the Bielefeld-Parma
collaboration put forward a method to compute finite-density 
QCD thermodynamic observables in the region to which access would be 
precluded by the sign problem; this approach is also able to probe
the singualrity structure of the theory in the complex $\mu_B$
plane \cite{schmidt_net-baryon_2021,singh_lee-yang_2022,nicotra_lee-yang_2022,Dimopoulos:2021vrk}. 
The method is based on the computation of Taylor series
coefficients at both zero and imaginary values of the baryonic
chemical potential, which enables the computation of multi-point 
Pad\'e approximants. This work aims to assess the effectiveness of the
method by making use of it in the context of a very standard
playground for the physics of phase transitions ({\em e.g.} the 2d
Ising model). At the same time, we present (very) preliminary results
on new applications in the context of finite-density QCD.\\

\begin{figure}[ht] 
\centering
\includegraphics[scale=0.62]{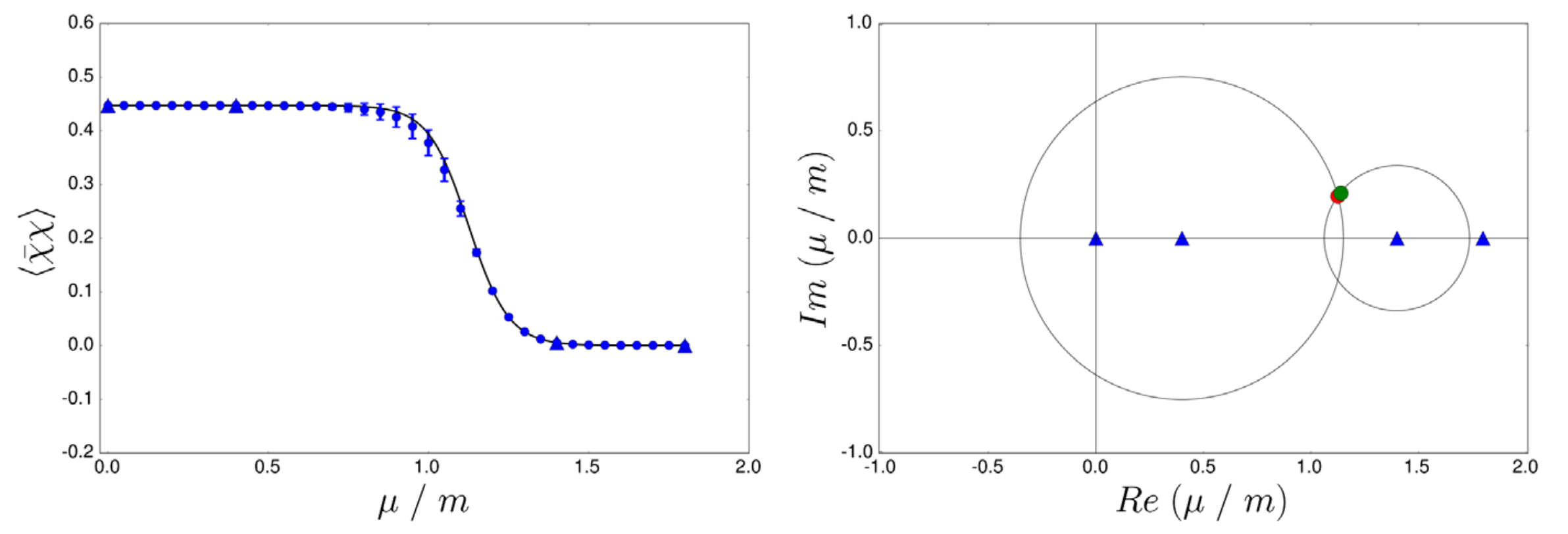}
\caption{Left panel: (continuum line) analytic solution for the
  condensate $\bar{\chi}\chi$ of the 1-D Thirring model 
($L=8$, $m=1$, $\beta=1$) at various values of the chemical potential
by mass ratio $\frac{\mu}{m}$; (triangles) numerical results obtained 
on one single thimble; (dots) numerical results taken from the rational
approximant. Right panel: we plot in the complex $\frac{\mu}{m}$ plane  
the singularity we got from the rational
approximant; it is depicted on top of the known analytic one.}
\label{fig:ThirringIDEA}
\end{figure}

Before entering the main subject, it is useful to recall when the idea
of applying multi-point Pad\'e rational approximants first came to our
mind; that was in the context of thimble regularisation. The latter
\cite{Cristoforetti:2012su,Fujii:2013sra} was introduced to solve (or at least tame)
the sign problem by re-expressing the path integral as a sum of
integrals computed on manifolds different from the original one. After
complexifying the degrees of freedom, one considers the so-called
Lefschetz thimbles, {\em i.e.} the manifolds that are the union of the
steepest ascent paths stemming from the various stationary points of
the action. On such manifolds the imaginary part of the action stays
constant, so that the sign problem reduces to the so-called {\em
  residual phase} which is there due to the Jacobian of the change of
variables. There is a thimble attached to each stationary point and
in principle all can give a contribution to the path integral. This is
referred to as the {\em thimble decomposition}. To make a long story
short, we recall that {\em (a)} not all the thimbles give a non-null
contribution, {\em (b)} this picture changes in different regions of
the parameters space of the theory ({\em i.e.} a given thimble can
contribute to the path integral in a region and not in another one)
and {\em (c)} there are cases in which a single thimble (usually the
so called {\em dominant} one, attached to the stationary point with
the lowest action) is enough to compute the answer one is interested
in. The latter observation gave raise to the {\em single thimble
  dominance} hypothesis, which was shown to hold in a few cases, but
failed in others. The first example of a failure was provided by the
1-D Thirring model \cite{Fujii:2015vha,Alexandru:2015sua}, where it
was clearly shown that a single thimble is not enough to account for
the known analytic result. It is nevertheless important to remark that
there are regions in which one single thimble is enough, and this was
the logical starting point for the success of a computation based on
multi-point Pad\'e rational approximants. The success of such 
approach \cite{DiRenzo:2020cgp} can be recognised 
in Fig.~\ref{fig:ThirringIDEA}. On the left, we display the known analytic
result for the chiral condensate $\bar{\chi}\chi$ of the 1-D Thirring model 
($L=8$, $m=1$, $\beta=1$) at various values of the chemical potential
by mass ratio $\frac{\mu}{m}$. This is plotted together with the
numerical results which we got: triangles are results
computed on one single thimble at points where we are able to show that
this is enough; dots are results taken from the multi-point
Pad\'e method that we will better describe in the next section. Here
it is enough to say that a few Taylor expansion coefficients were
computed at the points marked by triangles and from those the
multi-point Pad\'e approximant was computed. The right panel of the
figure shows how the singularity pattern of the solution was
reconstructed: the rational approximant displayed a singularity which
falls on top of the analytic one. Convergence radii of the Taylor
expansions we computed can be spotted, showing that there is an
intersection of convergence disks, validating the procedure of
bridging the two regions where we were able to compute single thimble
results: all in all, while the thimble decomposition is discontinuous,
the physical observable is not. The figure refers to a given choice of
lattice size, mass and $\beta$-value; we were able to show
\cite{DiRenzo:2021kcw} that the method can successfully account for
the extraction of the continuum limit. 

\section{Multi-point Padè method for finite density Lattice QCD}

\subsection{Basics of the multi-point Padè method}
\label{sec:basicMP}

Suppose we know a few Taylor expansion coefficients of a given
function $f(z)$ at different points $\{z_k\,|\,k=1 \ldots N\}$. The basic
idea of our multi-point Pad\'e approach is to approximate $f(z)$ by a
rational function $R^{m}_{n}(z)$, which we call a $[m/n]$ Pad\'e approximant
\begin{equation}
\label{eq:PadeRatFunct}
R^{m}_{n}(z) = \frac{P_m(z)}{\tilde{Q}_n(z)} = \frac{P_m(z)}{1+Q_n(z)} = \frac{\sum\limits_{i=0}^m \, a_i \, z^i}{1 + \sum\limits_{j=1}^n \, b_j \, z^j}\,.
\end{equation}
$R^{m}_{n}(z)$ ({\em i.e.} the $a_i,b_j$ coefficients defining it) can
be fixed by requiring that it reproduces the values of $f$ and a few
of its derivatives at the given points $\{z_k\}$. 
Provided that $n+m+1=Ns$ 
($f^{(s-1)}$ being the highest order derivative we
computed at each point), this is possible by requiring that
\begin{equation}
\label{eq:LinearProblem}
\begin{split}
 & \hdots \\ 
P_m(z_k) - f(z_k)Q_n(z_k) &= f(z_k) \\ 
P_m'(z_k) - f'(z_k)Q_n(z_k) - f(z_k)Q_n'(z_k) &= f'(z_k) \\ 
 & \hdots \\ 
\end{split}
\end{equation}
In Eq.~(\ref{eq:LinearProblem}) we only wrote $2$ out of $s$ equations
for $1$ out of $N$ points. It should be clear what the overall problem 
amounts to: we have to solve a linear system, the unknowns being the 
$\{a_i,b_j \,|\,i=1 \ldots m,\,j=1 \ldots n \}$. This is not the only
possible way to solve for $R^{m}_{n}(z)$, but
for the purpose of understanding our approach it suffices (the
interested reader can refer to \cite{Dimopoulos:2021vrk} for other
alternatives\footnote{Notice that this is the simplest
  setting also with respect to another point: there is no reason for 
strictly asking knowledge of the same number of derivatives at each point.}). 
It should be clear that 
\begin{itemize}
\item Not only $R^{m}_{n}(z)$ can reproduce our input pieces of
  information; by a natural {\em analytic continuation} it can {\em
    predict} values of $f$ in an extended region (to the extent we do
  not exit the region in which the approximation holds, which thing of
  course deserves care of its own): left panel of
  Fig.~\ref{fig:ThirringIDEA} is an example.
\item When a {\em zero} in the denominator of $R^{m}_{n}(z)$ is not
  canceled by a corresponding zero of the numerator, we face a {\em
    singularity} of the rational approximation, which is supposed to
  teach us something on the {\em singularity structure} of $f$; quite
  obviously, singularities live in the {\em complex} $z$ {\em plane}:
  right panel of Fig. \ref{fig:ThirringIDEA} is an example.
\end{itemize}

\subsection{First application of the multi-point Padè method to finite
  density LQCD}

\begin{figure}[ht]
	\centering
	\includegraphics[width=.50\textwidth]{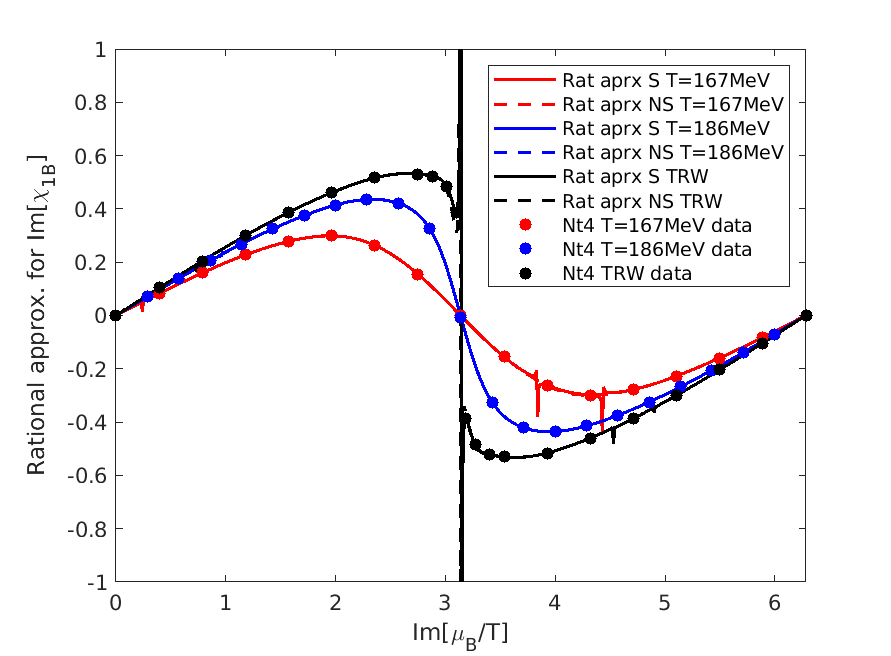}
	\hfill
	\includegraphics[width=.46\textwidth]{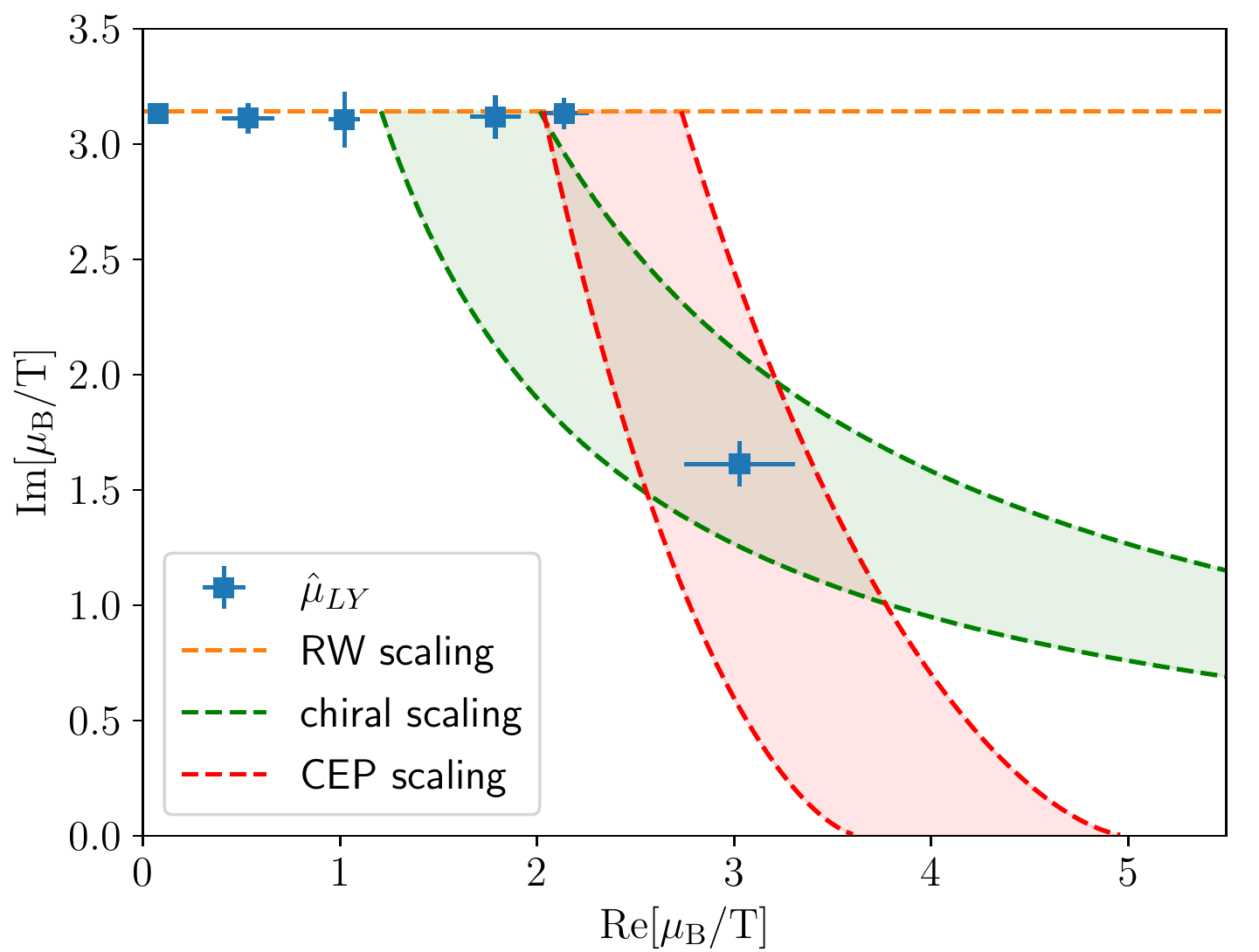}
	\caption{\label{fig:LQCD}(Left panel) The number density
          $\chi_{1B}$ at various values of $\hat\mu_B$ and different
          temperatures $T$. Data are shown together with two
          different rational approximants (enforcing parity or not):
          both describe data very well. The big spike is expected: it
        is the hint for the Roberge Weiss transition. (Right panel)
        The singularity pattern in the complex $\hat\mu_B$,
        highlighting their  expected overall compliance with Roberge
        Weiss, chiral and Critical End Point scaling.}
\end{figure}

In \cite{Dimopoulos:2021vrk} the Bielefeld Parma collaboration applied the
multi-point Padè method to finite density LQCD. In the example of
section~\ref{sec:PADEthimblePR} we did not have a way to safely
compute the 1D Thirring condensate in regions where more than one thimble give a
contribution; on the other hand, we could safely compute (on a single
thimble) at given values of $\frac{\mu}{m}$. This is the same as in LQCD: the sign
problem does not allow us to compute observables at {\em real} values of
the baryonic chemical potential $\mu_B$, but computations are safe at 
$\mu_B=0$ and at {\em imaginary} values of $\mu_B$ (in particular, we
can compute a few orders of the Taylor expansion of an observable). 
For (2+1)-flavor of highly improved staggered quarks (HISQ) 
\cite{Follana:2006rc} with imaginary chemical potential, we computed 
cumulants of the net baryon number density, given as 
\begin{equation}
    \chi_{nB}(T,V,\mu_B)=\left(\frac{\partial}{\partial \hat\mu_B}\right)^n\frac{\ln Z(T,V,\mu_l,\mu_s)}{VT^3},
\end{equation}
with $\hat\mu_B=\mu_B/T$ and $l,s$ referring to light and strange
flavors. Dependence on masses is not made explicit: the light to
strange ratio is the physical one. By computing at different imaginary
values of  $\hat\mu_B$ (including $\hat{\mu}_B=0$) we could implement the program
of subsection~\ref{sec:basicMP}. Fig.~\ref{fig:LQCD} is the counterpart
of Fig.~\ref{fig:ThirringIDEA}. 
We point out that
\begin{itemize}
\item In the left panel we can see how well the rational approximants
  for the number density $\chi_{1B}$ describe data at different
  temperatures. Actually we show two different
  rational approximants (enforcing parity or not): they are both
  fine. The big spike is expected to be there: it is related to the
  Roberge Weiss transition, and it occurs at the temperature which is
  supposed to be the relevant one ($T_{RW}$). Minor spikes can be also
  spotted: they are harmless, and they can be understood in terms of
  what we will explain in the next section (partial cancellation of
  zeros between numerator and denominator). 
\item In the right panel we display the singularities we found at
  different temperatures, relating them to the expected singularity
  scaling pattern. These are the expected Lee-Yang singularities: 
  one expects a given scaling for the singularities
  connected to the Roberge Weiss transition, to the chiral transition
  and to the QCD Critical End Point. While the last two are still
  under investigation\footnote{Indeed we now have an estimate for the
    CEP Temperature.}, one can clearly see a consistent picture for
  the Roberge Weiss scaling: indeed in \cite{Dimopoulos:2021vrk} we were able to
  show that it is the expected one.
\end{itemize}
All in all, results are intriguing. That's why we now want to show
that the machinery is under control for the the most popular
  arena for testing tools in the study of phase transitions, {\em i.e.}
the two-dimensional Ising model.

\section{Testing the method on the 2d Ising model}

\begin{figure}[ht]
	\centering
	\includegraphics[width=.495\textwidth]{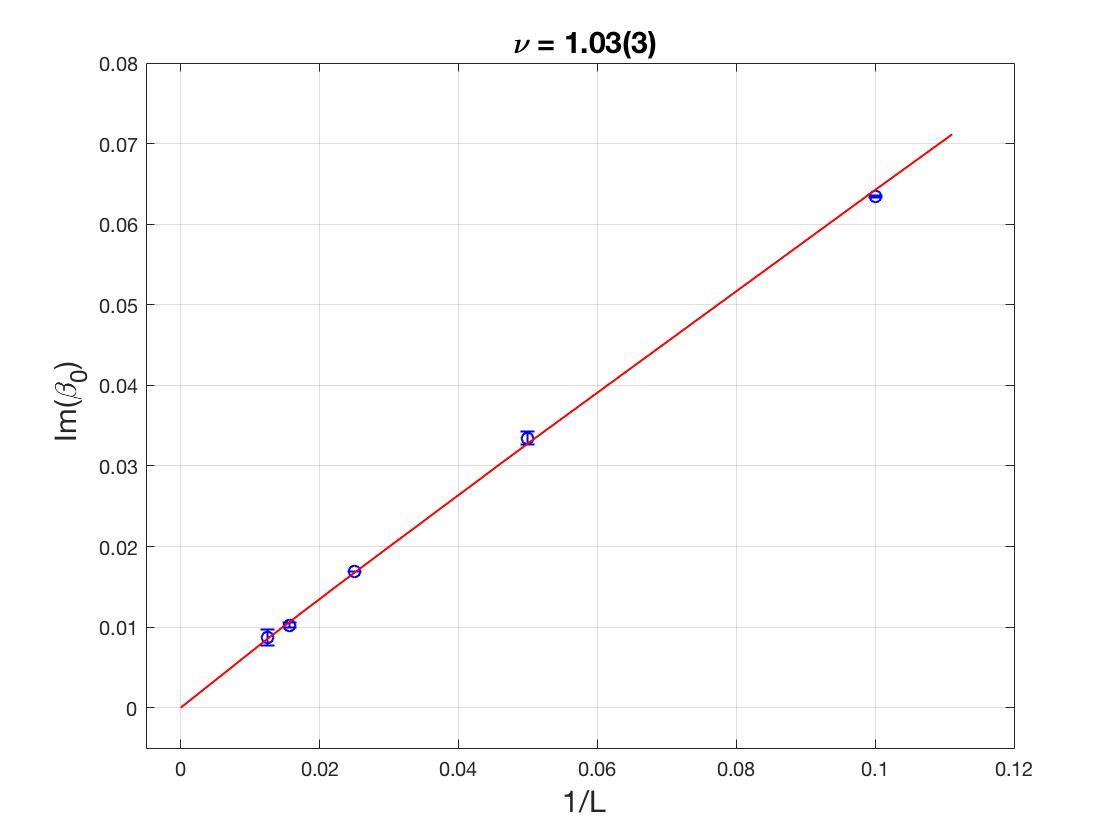}
	\hfill
	\includegraphics[width=.495\textwidth]{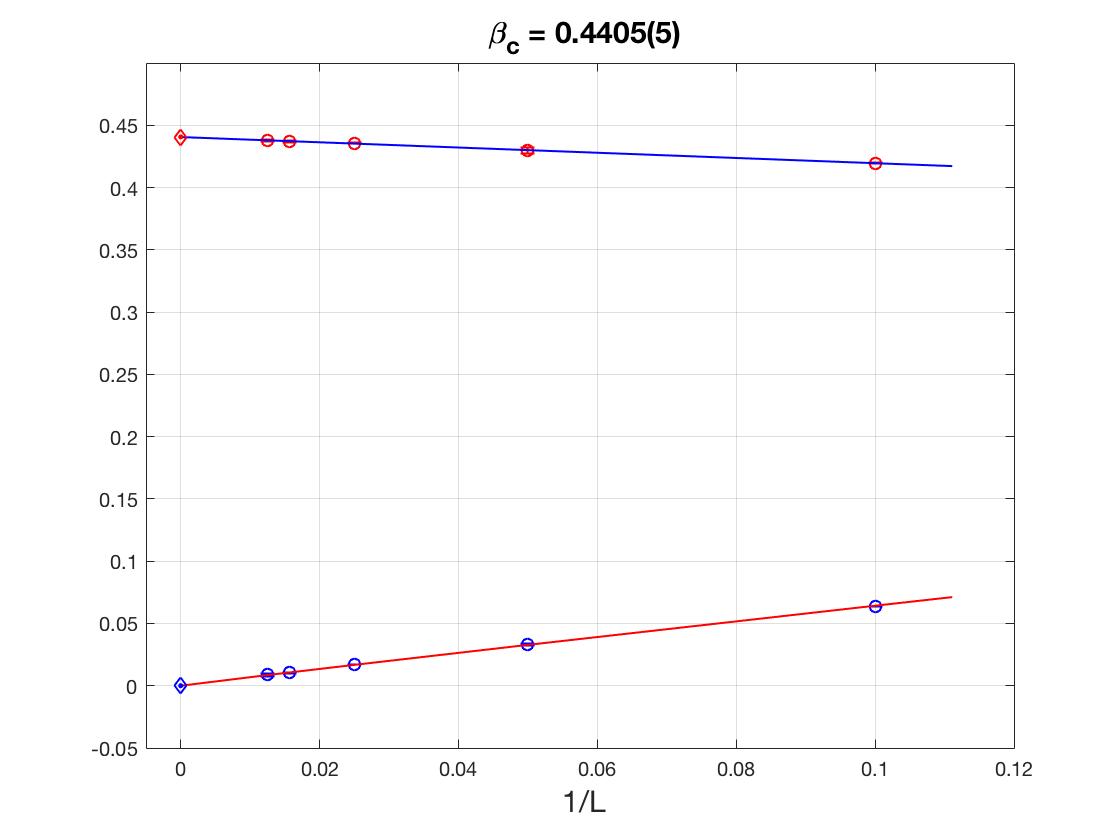}
	\caption{\label{fig:FisherZeros}(Left panel) The scaling in
          $1/L$ of $\mbox{Im}(\beta_0)$, {\em i.e.} 
          the imaginary part of the Fisher zero, detected as that the
          closest singularity of the cumulant to
          the real axis. The correct critical exponent $\nu=1$ is got
          with fairly good accuracy. (Right panel) Once $\nu$ has been
          recognised to be the right one, one can fit the value of the
          critical inverse temperature $\beta_c$, which is
          reconstructed to per mille accuracy.}
\end{figure}

\begin{figure}[b] 
\centering
\includegraphics[scale=0.62]{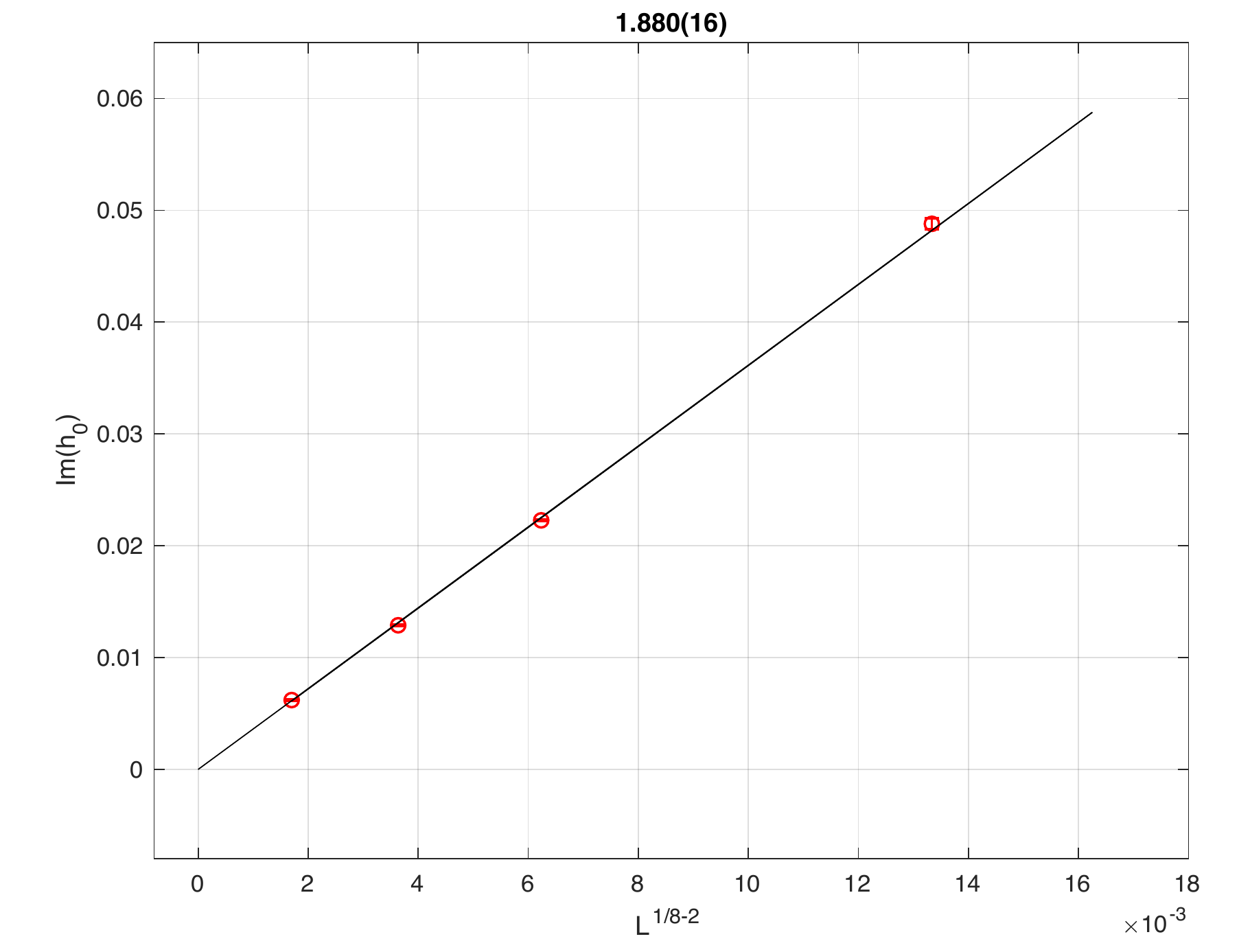}
\caption{Finite size scaling of $\mbox{Im}(h_0)$. To guide the eye, we
plot data versus $L^{1/8-2}$, where the correct critical exponent is
taken. As the figure title we report the absolute value of the one we
got, which turns out to be a very accurate estimate, to less than percent.}
\label{fig:LeeYangScaling}
\end{figure}

Lee-Yang theory is one of the possible approach to the study of phase
transitions. For an example of its application, we refer the
interested reader to \cite{Deger:2019mgo}, where the authors study the
2d Ising model. We will basically follow their program, but will not
rely on the study of many different cumulants (as they do). We will instead make use
of our multi-point Padè method and study only two different cumulants at
different values of temperature and magnetic field. The hamiltonian is
the well-known one, based on interactions between nearest neighbours and
with external magnetic field $h$
\begin{equation}
H = - J \sum_{<i,j>} \sigma_i \sigma_j - h \sum_i \sigma_i
\end{equation}
with the only possible values $\sigma_i=\pm 1$. In the following $J$
will be set to $J=1$. The partition function can be written in terms
of its {\em zeros} $\{\beta_k\}$
\begin{equation}
Z(\beta,h) = Z(0,h) \, e^{\,\beta c} \, \prod_k (1-\frac{\beta}{\beta_k})
\end{equation}
$c$ being a constant. If we define thermal cumulants by
$$ \left\langle\langle U^n \right\rangle\rangle =
\frac{\partial^n}{\partial (-\beta)^n} \ln Z(\beta,h)$$
it is easy to show that they can be expressed as
\begin{equation}
\left\langle\langle U^n \right\rangle\rangle =
(-1)^{(n-1)} \sum_k \frac{(n-1)!}{(\beta_k-\beta)^n} \;\;\;\;\;\;\; (n>1)
\end{equation}
Furthermore, scaling relations describe the approach of leading zeros to critical inverse temperature
\begin{equation}
\label{eq:scalingZeros}
| \beta_0-\beta_c | \sim L^{-1/\nu} \;\;\;\;\;\;\; \mbox{Im}(\beta_0)  \sim L^{-1/\nu}.
\end{equation}
In Eq.~(\ref{eq:scalingZeros}) $\beta_0$ is the Fisher zero, that is the
closest zero of the partition function to the real axis, resulting in
the closest singularity of cumulants to the real axis\footnote{$\beta_0$ shows up together with
  its complex conjugate $\beta^*_0$.}, $\beta_c$ is the critical inverse
temperature and $\nu$ is the relevant critical exponent. \\
Our program now entails four steps: {\em (1)} we compute the $n=2$
thermal cumulant ({\em i.e.} the specific heat) at
  various inverse temperatures $\beta$ and lattice sizes $L$; {\em (2)} for each $L$ we compute the rational approximant
  $R^{m}_{n}(\beta)$ by our multi-point Padè method; {\em (3)} at each
  $L$ we find the Fisher zero $\beta_0$, which is obtained as the
  the closest singularity of the cumulant to the real axis; {\em (4)} we study the finite size scaling
  of the values of $\beta_0$. The result of the procedure can be
  inspected in Fig.~\ref{fig:FisherZeros}.
\begin{itemize}
\item In the left panel we display the scaling in $1/L$ of
  $\mbox{Im}(\beta_0)$. Errors are computed by varying results with respect to
  statistical errors for the cumulant and functional form for the
  rational approximant. As one can see, the value of the relevant
  critical exponent $\nu=1$ is got with fairly good accuracy ($1.03(3)$).
\item Once $\nu=1$ has been recognised, we can fit the scaling of the
  real part $\mbox{Re}(\beta_0)$ (right panel), thus finding the value
  of the critical inverse temperature. We get the very accurate result
  $\beta_c=0.4405(5)$. 
\end{itemize}
Once the critical inverse temperature is known, one can sit on top of
it and study the scaling in $L$ of $\mbox{Im}(h_0)$, $h_0$ being the
Lee Yang zero, that is the closest singularity of a magnetic 
cumulant to the real axis. Explicitly,  
our program again entails four steps: {\em (1)} we compute the $n=1$
magnetic cumulant ({\em i.e.} the magnetisation) at
  $\beta=\beta_c$ and various values of external magnetic field $h$ and 
  lattice size $L$; {\em (2)} for each $L$ we compute the rational approximant
  $R^{m}_{n}(h)$ for the magnetisation by our multi-point Padè method; {\em (3)} at each $L$
  we find the Lee Yang zero $h_0$, which is the singularity of the
  rational approximant for the  magnetisation which is the closest 
to the real axis; {\em (4)} we study the finite size scaling
  of the values of $\mbox{Im}(h_0)$ (as we will see, $h_0$ always sits
  at $\mbox{Re}(h_0)=0$). \\
\begin{figure}[b]
	\centering
	\includegraphics[width=.5\textwidth]{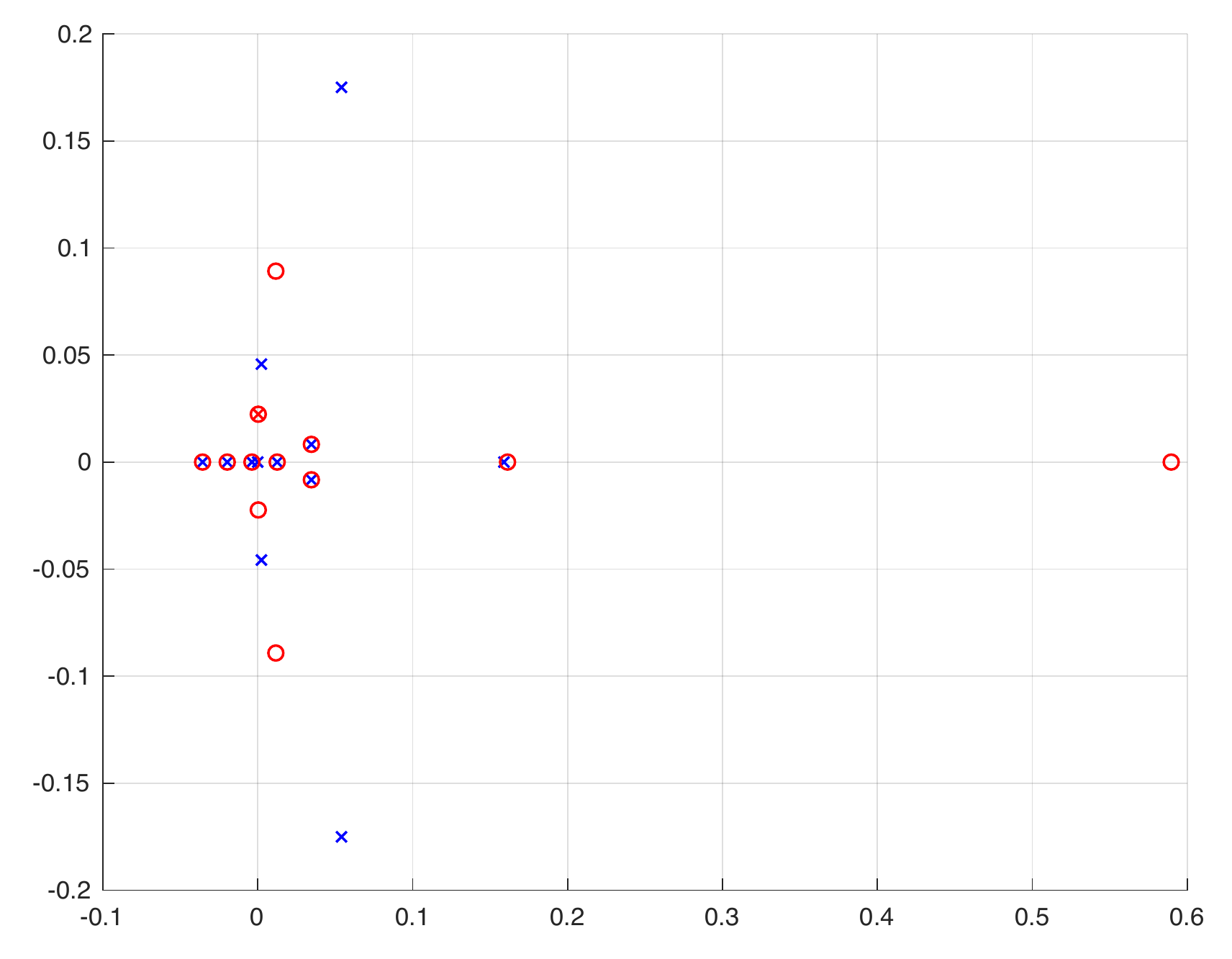}
	\hfill
	\includegraphics[width=.49\textwidth]{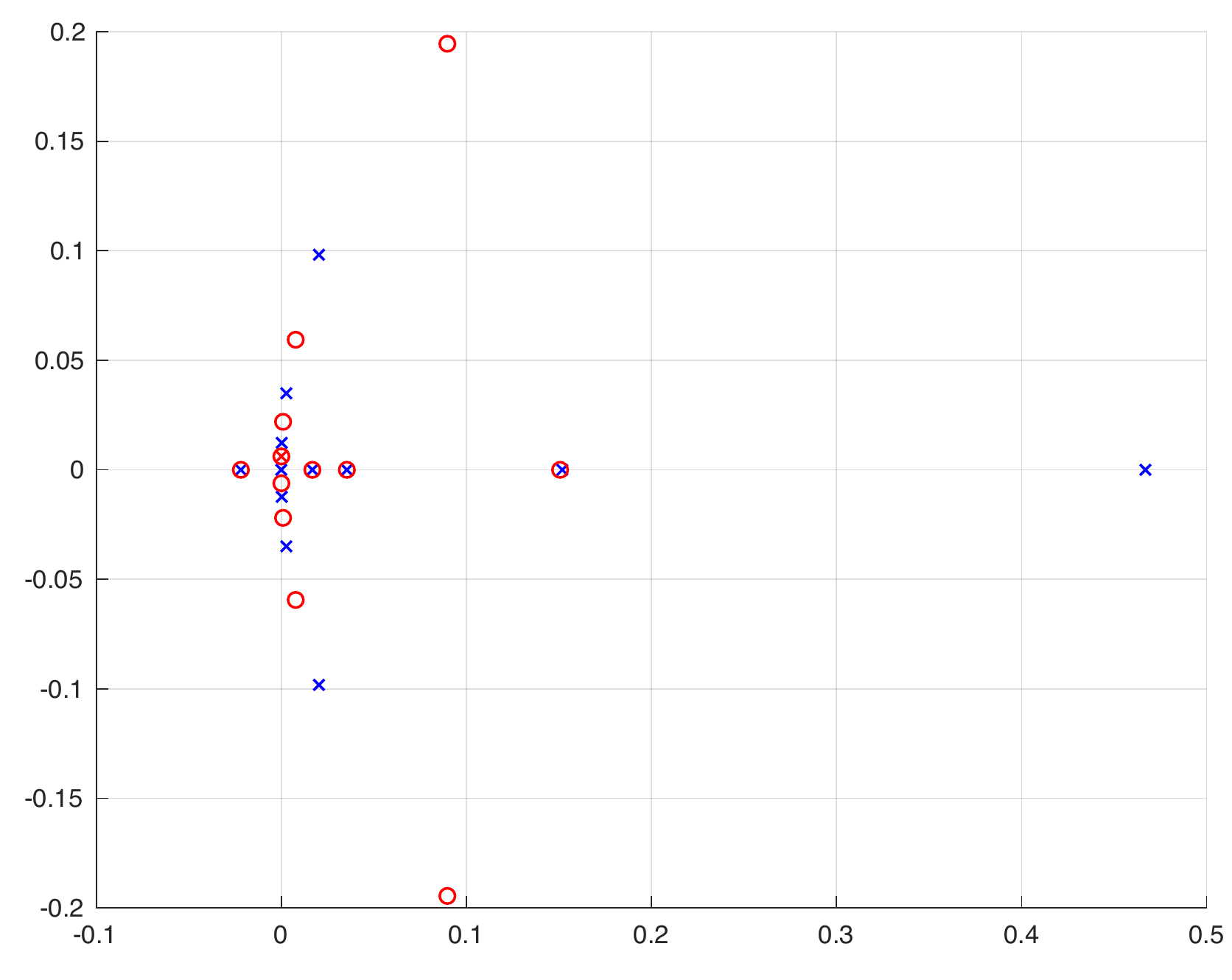}
	\caption{\label{fig:LeeYangZerosL}(Left panel) Zeros of
          the numerator (blue crosses) and of the denominator (red
          circles) of the rational approximant $R^{m}_{n}(h)$ for the
          magnetisation on $L=15$ (left panel) and $L=30$ (right
          panel). We highlight the closest singularity to the real
          axis, which is getting closer to the real axis itself as $L$
        gets larger, with real parts being $\mbox{Re}(h_0) =
        0$. Plots are in the complex $h$ plane.}
\end{figure}
\noindent
Before we inspect this scaling behaviour, it
  is useful to have a closer look at the singularity pattern  in the
  complex $h$ plane at given
  values of $L$. In Fig~\ref{fig:LeeYangZerosL} we depict the zeros of
  the numerator (blue crosses) and of the denominator (red circles) of our
  $R^{m}_{n}(h)$ at different values of the lattice size $L$, {\em
    i.e.} $L=15$ (left panel) and $L=30$ (right panel).
We can easily make a couple of key observations. 
\begin{itemize}
\item A few zeros of the denominator are canceled by corresponding zeros of
  the numerator. These are not genuine pieces of information: actually
  their location vary when varying {\em e.g.} the order of the Pad\'e
  approximant $[m,n]$. On the other hand, genuine pieces of information ({\em i.e.}
  actual zeros and poles) stay constant to a very good
  precision. Notice that this is the explanation for the small spikes
  in Fig.~\ref{fig:LQCD}: they are simply the shadow of cancellations
  which are indeed very good, but not good enough to be invisible when
  plotting the rational approximant.
\item We can clearly see that, as the lattice size $L$ gets larger,
  the closest singularity (Lee Yang zero, highlighted in the plot)
  gets closer to the real axis, with real parts being $\mbox{Re}(h_0)
  = 0$.
\end{itemize}
Finally, in Fig.~\ref{fig:LeeYangScaling} we plot the finite size scaling
  of $\mbox{Im}(h_0)$. As one can see, the critical
  exponent in is got with very good accuracy (this time, less than
  percent: $-1.880(16)$ vs $-1.875$).
The steps we could take in the (much simpler) case of the
Ising model would be the preferred conceptual path to follow also for LQCD. 
Needless to say, it will take time before we can be in a position to
do that.

\section{Back to LQCD: a T-Pad\'e application}

\begin{figure}[t]
	\centering
	\includegraphics[width=.495\textwidth]{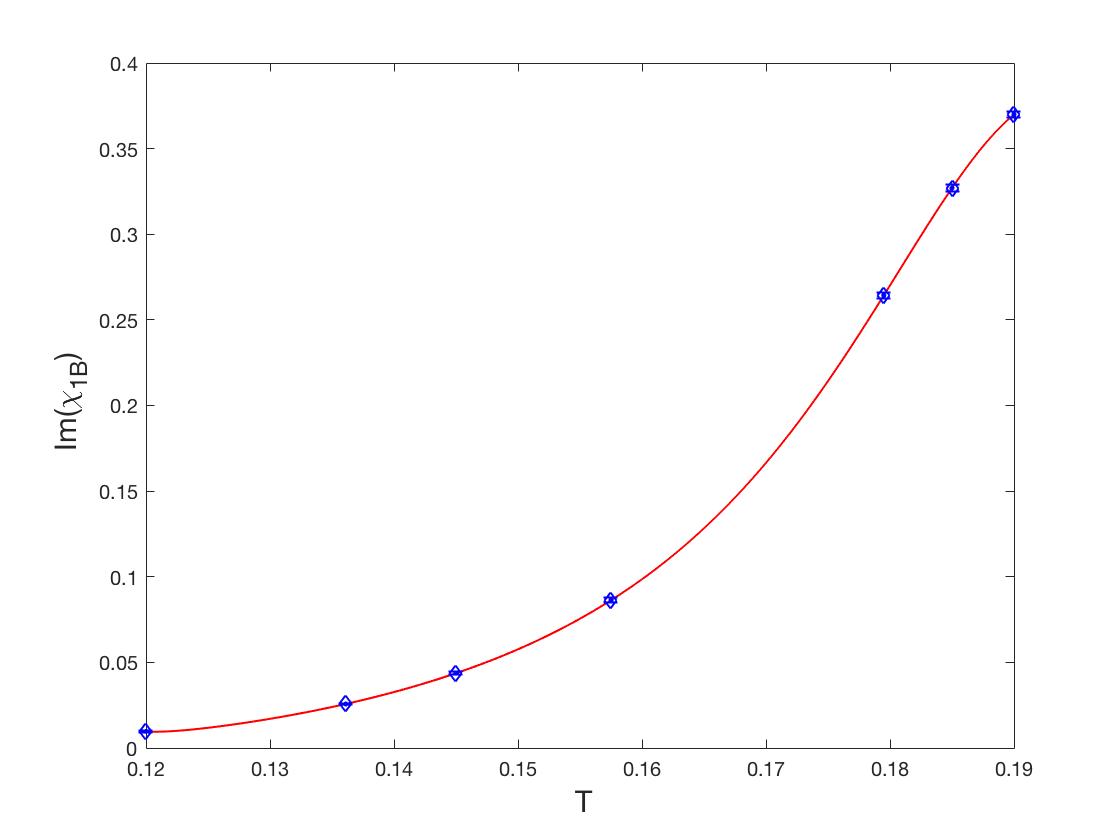}
	\hfill
	\includegraphics[width=.495\textwidth]{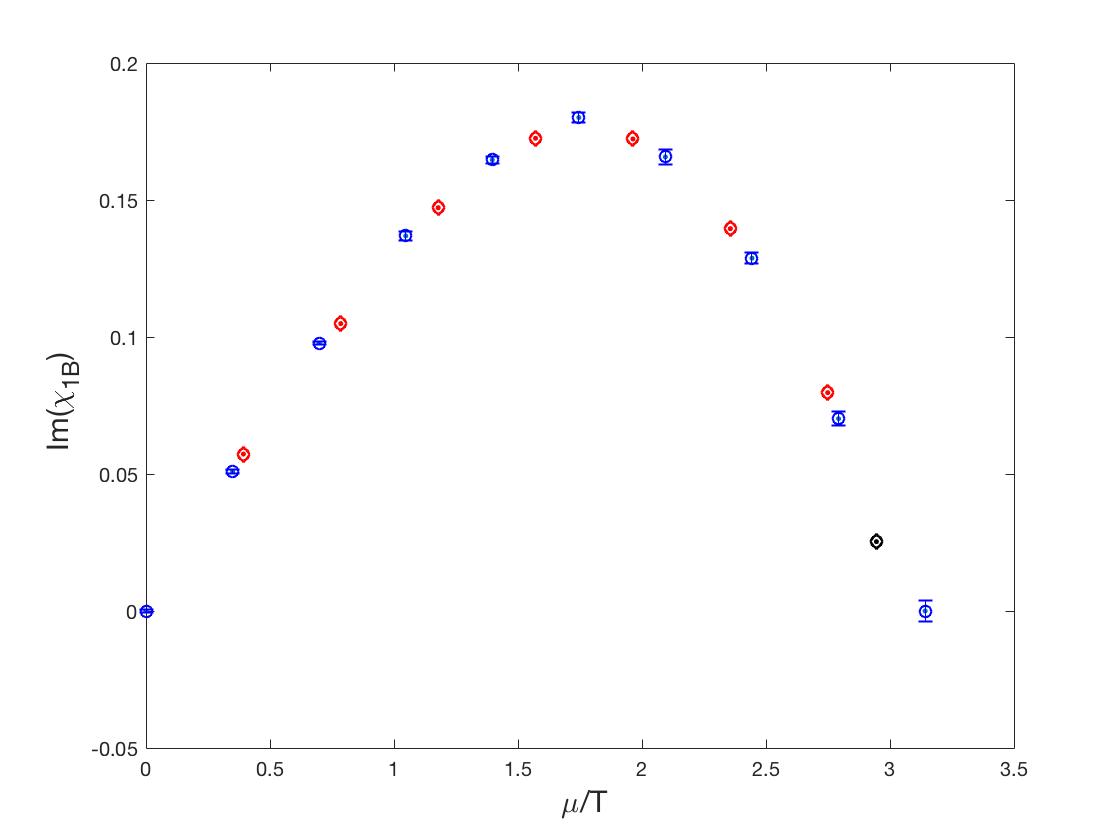}
	\includegraphics[width=.495\textwidth]{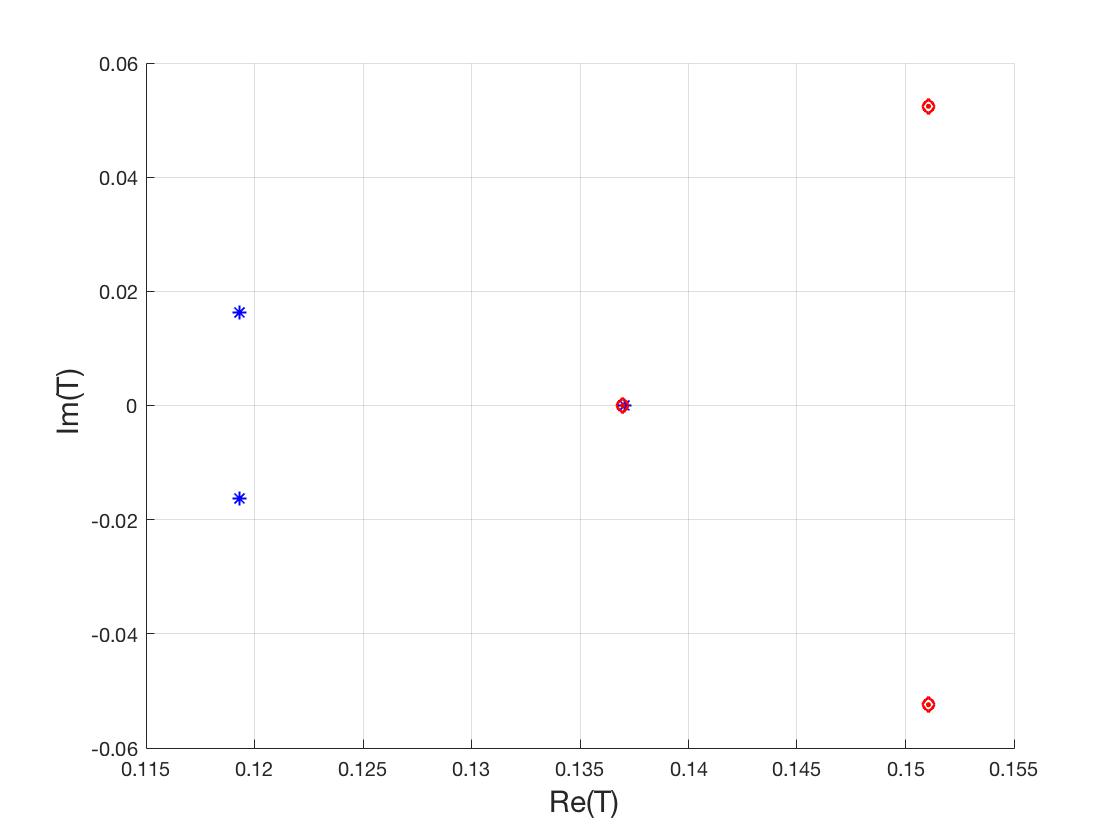}
	\hfill
	\includegraphics[width=.495\textwidth]{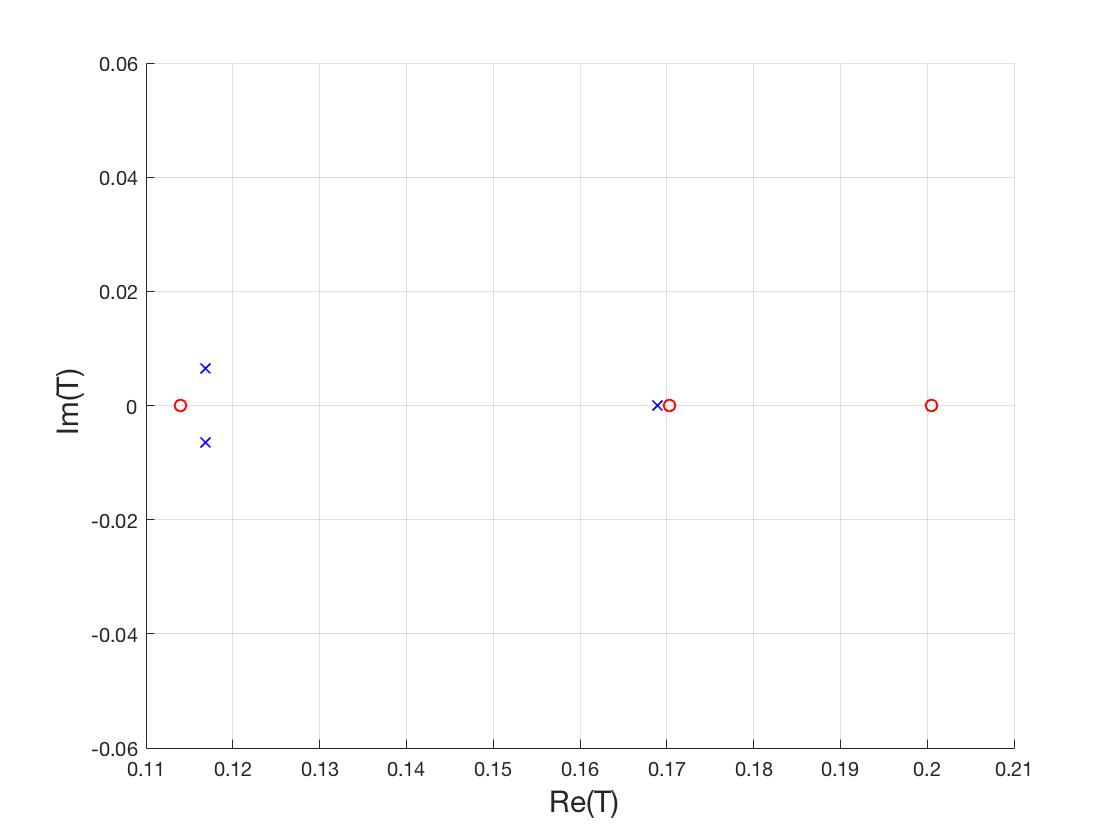}
	\caption{\label{fig:PadeT}(Top-left panel) An example of 
$R^{m}_{n}(T)$ for $\chi_{1B}$ at a given value of $\hat{\mu}_B$ on top of data taken at different
temperatures $T$ at the same given value of $\hat{\mu}_B$. (Top-right) Actual measurements
of $\chi_{1B}(\hat{\mu}_B)$ at a given temperature $T$ plotted
together with interpolating data obtained from
$R^{m}_{n}(T)$. Everything looks pretty smooth; we plot in a different
colour the only data point possibly not falling smoothly on top of
actual data. (Bottom-left) Zeros of denominator (red) and zeros of numerator (blue) of $R^{m}_{n}(T)$ in the complex $T$ plane
at a low value of $\hat{\mu}_B$.
(Bottom-right) The same plot 
at a value of $\hat{\mu}_B$ close to $\hat{\mu}_B=i \pi$ ($T$ is expressed in GeV) 
}
\end{figure}

We finally go back to LQCD for a (very) preliminary account of a new
application. Till now we have seen multi-point Padè approximants from
data taken at a given temperature $T$ and different values of
$\hat{\mu}_B$: with this we mean that we obtained different
$R^{m}_{n}(\hat{\mu}_B)$ at different $T$ values. 
With the very same data, we can think of going the other
way around, that is we can obtain $R^{m}_{n}(T)$ at different
$\hat{\mu}_B$ values. Fig.~\ref{fig:PadeT} is an example of what we
can get following this path. Of course, this time singularities
emerge in the complex $T$ plane.

\section{Conclusions}

The multi-point Padè method for the study of phase transitions has
already proved to be quite effective in the case of LQCD. 
Here we showed how the approach can provide very accurate results 
when collecting a rich statistics is not such a hard numerical task
(as it was the case for the 2d Ising model). This is at same time a
proof of concept of the reliability of the method and a stimulus to do
better in the case of finite density LQCD.

\section*{Acknowledgements}

This work has received funding from the European Union's Horizon 2020
research and innovation programme under the Marie Sk{\l}odowska-Curie
grant agreement No. 813942 (EuroPLEx). We also acknowledge support from I.N.F.N.
under the research project {\sl i.s. QCDLAT}. This work benefits from
the HPC facility of the University of Parma, Italy.

\bibliographystyle{JHEP}
\bibliography{my_bib}

\end{document}